\documentclass[a4paper]{article}

\usepackage{INTERSPEECH2019}
\usepackage{multirow}
\usepackage{textcomp}

\title{Audio Spoofing Verification using Deep Convolutional Neural Networks by Transfer Learning}
\name{Rahul T P$^1$, P R Aravind$^1$, Ranjith C$^1$, Usamath Nechiyil$^1$, Nandakumar Paramparambath$^1$}

\address{
  $^1$Department of Electronics and Communication, NSS College of Engineering, Palakkad, India}
\email{rahultp.rtp@gmail.com, aravindpkd1@gmail.com, ranjithcotp97@gmail.com, usmathn@gmail.com, nandakumarpp@hotmail.com}

\begin{document}

\maketitle

\begin{abstract}
    Automatic Speaker Verification systems are gaining popularity these days; spoofing attacks are of prime concern as they make these systems vulnerable. Some spoofing attacks like Replay attacks are easier to implement but are very hard to detect thus creating the need for suitable countermeasures. In this paper, we propose a speech classifier based on deep-convolutional neural network to detect spoofing attacks. Our proposed methodology uses acoustic time-frequency representation of power spectral densities on Mel frequency scale (Mel-spectrogram), via deep residual learning (an adaptation of ResNet-34 architecture). Using a single model system, we have achieved an equal error rate (EER) of 0.9056\% on the development and 5.32\% on the evaluation dataset of logical access scenario and an equal error rate (EER) of 5.87\% on the development and 5.74\% on the evaluation dataset of physical access scenario of ASVspoof 2019.
 
\end{abstract}


\section{Introduction}

    Biometrics technologies play a critical role in regulating access to informational resources in today\textquotesingle s world. One of the reliable approaches for attaining a suitable secure system is speaker verification.
    
    Automatic speaker verification (ASV) has undergone rapid improvements in the recent decade but continues to show high vulnerability towards different spoofing attacks. Spoofing methods are categorized as speech synthesis (SS), voice conversion (VC), impersonation and replay attacks \cite{spoofingmethods}. Among these, replay attacks are arguably the most simple yet highly indistinguishable ASV spoofing technique as they do not require the attackers to have any specialized knowledge, and can be mounted with relative ease using consumer devices.
    
    For the first automatic speaker verification spoofing and counter-measures challenge (ASVspoof 2015 \cite{asvspoof15}) even though the best results showed an overall average detection EER of less than 1.5\%, the EER of unknown attacks is five times higher than that of known attacks. In addition, while some attacks were easily and consistently detected, others provoked extremely high error rates nearing 50\%.
    
    Systems submitted under ASVspoof 2017 \cite{asvspoof17} challenge inspected distinct front-end features and the usage of various classifiers to detect replay attack under conditions. Among them, the best-performing system \cite{paper4}, had an equal error rate (EER) of 6.73\% and used a light convolutional neural network (LCNN) to extract high-level features from the log power spectrum, together with a Gaussian Mixture Model (GMM) as classifier. Variable length Teager energy operator-energy separation algorithm-instantaneous frequency cosine coefficients (VESA-IFCCs) \cite{paper5} were proposed as a single system with the aim of capturing the spectral changes due to the transmission and channel characteristic of replay devices. A single frequency filtering feature with a high spectro-temporal resolution was proposed \cite{paper6} to capture the channel information embedded in the low signal to noise ratio region.%1473.pdf
    
    ASVspoof 2019 \cite{asvspoof-org} Challenge was organized to emphasize the development of reliable countermeasures that could efficiently segregate bona fide and spoofed speech. The initiative aims specifically to encourage the design of generalized countermeasures, i.e. countermeasures that would perform well when encountered with spoofing attacks which are unpredictable in nature. The data used for ASVspoof 2015 \cite{asvspoof15} contained spoofed speech samples generated using text-to-speech (TTS) and voice conversion (VC) systems at that time. The ASVspoof 2017 \cite{asvspoof17} challenge emphasized on the design of countermeasures strived at detecting replay spoofing attacks that could be implemented easily by anyone using conventional devices. The ASVspoof 2019 \cite{asvspoof-org} edition was the first to focus on countermeasures for all three major spoofing attack types, namely those originating from TTS, VC and replay attacks. ASVspoof 2019 \cite{asvspoof-org} focuses to develop next-generation countermeasures for the automatic detection of spoofed or fake audio. The challenge encloses two separate sub-challenges in logical and physical access control and provides a common database of the most advanced spoofing attacks to date. The aim was to study the extremities of spoofing countermeasures with respect to automatic speaker verification and spoofed audio detection.
    
    In this paper, we propose the use of Mel-spectrograms \cite{melref} which is obtained from the audio frames as a time-frequency representation of power spectral density, for the training of deep-convolutional neural networks for audio spoofing attack detection. The use of Mel-spectrograms provides a time-frequency representation with sufficient spectral resolution, which is required for robust replay attack detection. Our framework is based on adapting the ResNet-34 architecture \cite{resnet}.

\section{Methodology}
\begin{figure}[t]
  \centering
  \includegraphics[width=\linewidth]{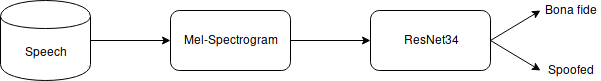}
  \caption{Overview of the proposed framework for speech spoofing detection.}
  \label{fig:methodology}
\end{figure}

    Our proposed framework (Figure~\ref{fig:methodology}) employs transfer learning \cite{transref} of a pre-trained convolutional neural network (CNN) for fast adaptation to the Mel-spectrograms extracted from speech inputs. Transfer learning is an analysis technique in machine learning that used to solve a problem by reusing knowledge gained from solving another but similar problem. Extracted spectrograms are fed into the deep convolutional neural network in which the speech signals are classified as bona fide or spoofed. In the following subsections, we describe the three major components of our system: Transfer Learning \cite{transref}, Mel-spectrograms \cite{melref} and ResNet \cite{resnet}, followed by a functional overview of our proposed framework. 

\subsection{Transfer Learning}
    Transfer Learning is a machine learning technique that is used to solve a problem based on knowledge gained from solving another but related problem. In this approach, the features gained from training a base neural network is reused or adjusted to a target neural network for its training. Transfer learning is usually expressed through the use of pre-trained models. A pre-trained model is trained using a large benchmark dataset to solve a similar problem. The computational cost of training such models is high, So it is common practice to reuse models from published literature (e.g. VGG, Inception, MobileNet, ResNet). Pre-trained models used in transfer learning are usually based on large convolutional neural networks (CNN). High performance and the easiness in training are two of the main factors that makes CNN popular for such applications.

\subsection{Mel-spectrograms}

\begin{figure}[t] 
  \centering
  \includegraphics[width=\linewidth]{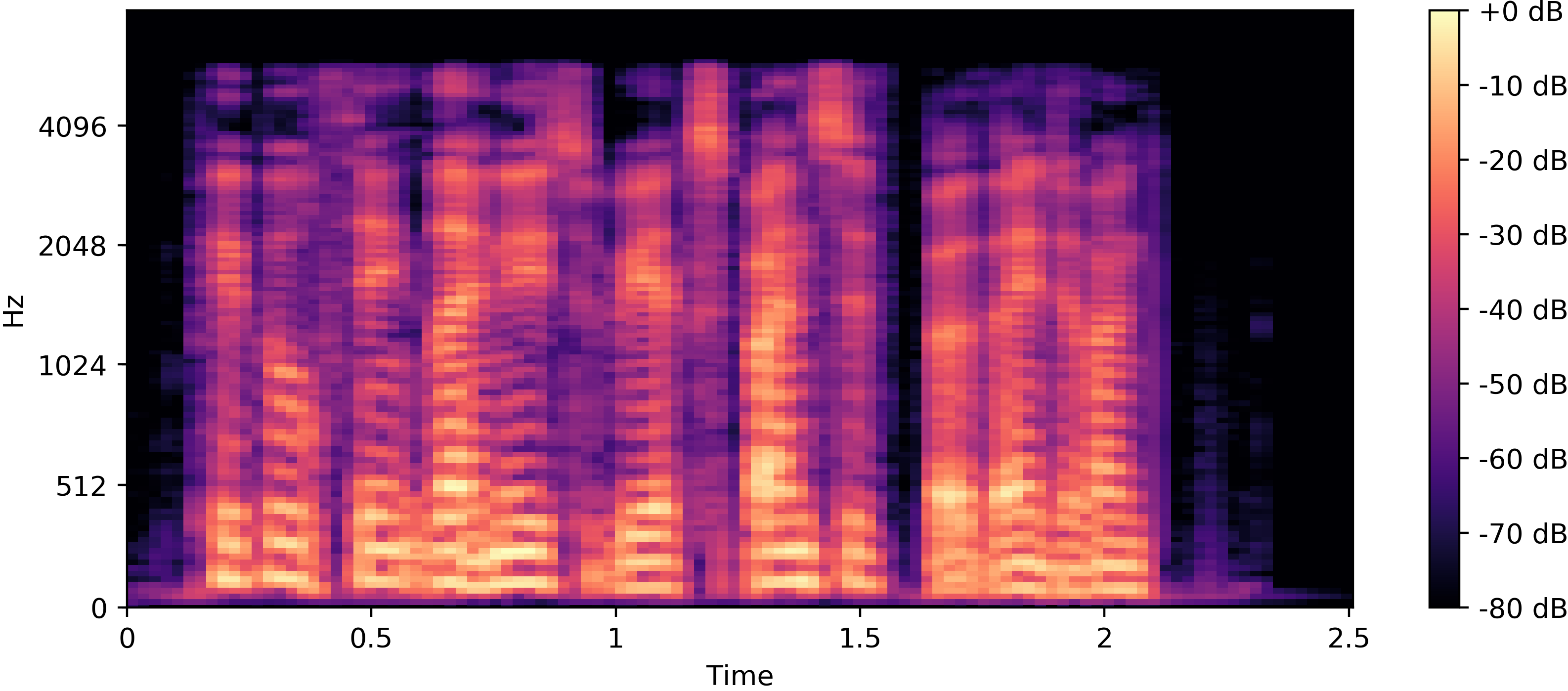}
  \caption{Mel-spectrogram of an audio.}
  \label{fig:melspectrogram}
\end{figure}

    A Mel-spectrogram represents an acoustic time-frequency representation of a sound, i.e. the power spectral density $P(f, t)$. It is sampled into a number of points at equal intervals of time $t\textsubscript{i}$ and frequencies $f\textsubscript{j}$ (on a Mel frequency scale).
    
    The Mel-frequency scale is defined as:
        \begin{equation}
            Mel = 2595 log\textsubscript{10}(1 + \frac{Hertz}{700})
        \end{equation}
    And its inverse is:
    	\begin{equation}
            Hertz = 700 (10\textsuperscript{$\frac{Mel}{2595}$} - 1)
        \end{equation}
    %http://www.fon.hum.uva.nl/praat/manual/MelSpectrogram.html
    STFT (short-time Fourier transform) and Mel-spectrogram have been the most popular input representations for audio classifications \cite{melref}. Mel-spectrograms provide an efficient and perceptually relevant representation compared to STFT \cite{hearref} and have been shown to perform well in various tasks \cite{melperf1}-\cite{melperf5}. However, an STFT is closer to the original signal and neural networks may be able to learn a representation that is more optimal to the given task than Mel-spectrograms. The latter requires large amount of training data; however as reported in \cite{melexp}  where using Mel-spectrograms outperformed STFT with a smaller dataset.
    % REFER CHOI2018

   Mel-spectrogram transforms the input raw sequence to a 2-D feature map where one of the dimensions represents time and the other one represents frequency and the values represent amplitude. The above Figure \ref{fig:melspectrogram} shows the Mel-spectrogram of an audio signal. 2-D Mel Model betters the 1-D raw wave model but the average of the two outperforms each individual model significantly \cite{1d2d}. The Mel-spectrogram is selected since it is psychologically relevant and computationally efficient. It produces a Mel-scaled frequency representation which is a more appropriate conjecture of human auditory understanding \cite{hearref} and typically involves compressing the frequency axis of short-time Fourier transform representation. And the magnitude of Mel-spectrogram is mapped to the decibel scale.
   % REFER 1703.09179

\subsection{Deep Residual Net (ResNet)}
    The deeper the Neural Network the more difficult it is to train. It is well known that the depth of a network is a determining factor of the network performance. Deeper neural networks are used in the field of computer vision. However, it is not easy to train a deeper network due to the notorious gradient vanishing problem. This was one of the drawbacks of VGG net \cite{vggnet} as they started losing generalization capability after some depth. To address this problem, an effective method called residual neural network(ResNet)\cite{resnet} is introduced. ResNet provides a training framework to ease the training of networks that are substantially deeper than those used previously. It was motivated by the counter intuitive experimental findings that adding more layers lead to higher training error. Theoretically, as the number of layers increased, the modeling capabilities of the Neural Networks should be better and therefore, the deeper networks should produce no higher training error. Experiments in \cite{resnet} showed that ResNet greatly improves training efficiency since the gradients can propagate many layers through the shortcut connection. In addition, ResNet allows deeper networks to be trained, resulting in models that usually perform better.
   %refer 1085 pdf full nokkanam resnet.
    One of the problems ResNets solves is known as the vanishing gradient. This is because when the network is too deep, the gradients from where the loss function is calculated easily shrink to zero after several applications of the chain rule. This results in the weights never getting updated and therefore, no learning process is involved. With ResNets, the gradients flow directly through the identity function.
    
\begin{figure}[t]
  \centering
  \includegraphics[width=0.5\linewidth]{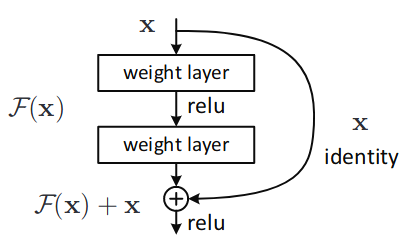}
  \caption{Residual learning: a building block \cite{resnet}.}
  \label{fig:residualblock}
\end{figure}

    Deep residual learning \cite{resnet} enables the training of CNNs that have substantially deeper architectures. It introduces skip connections that enable gradient flow across a large number of layers thus relieves the problem of vanishing gradients in deep CNNs. These skip connections enables the outputs to learn a residual mapping. The residual block forms the basic building block of a ResNet (Figure \ref{fig:residualblock}). If the desired mapping to be learned is $\mathcal{H}(x)$, the stacked residual layers learn the residual mapping, $\mathcal{F}(x) = \mathcal{H}(x) - x$. Thus, the original mapping to be learned becomes $\mathcal{F}(x) + x$.  ResNet uses the Rectified linear unit (ReLU) activation function.
    
\section{Experiments}

\subsection{Dataset}

\begin{table}[ht]
\caption{Number  of  non-overlapping  target  speakers  and  number  of  utterances  in  training  and development sets of the ASVspoof 2019 database \cite{evaluation}.}
\label{tab:dataset}
\centering
\resizebox{\linewidth}{!}{%
\begin{tabular}{ccccccc}
\toprule
                        & \multicolumn{2}{c}{Speakers}                    & \multicolumn{4}{c}{Utterances}                                           \\ \cmidrule(lr){2-3} \cmidrule(l){4-7}
\multirow{2}{*}{Subset} & \multirow{2}{*}{Male} & \multirow{2}{*}{Female} & \multicolumn{2}{c}{Logical Access} & \multicolumn{2}{c}{Physical Access} \\ \cmidrule(lr){4-5} \cmidrule(l){6-7}
                        &                       &                         & Bona fide         & Spoofed        & Bona fide         & Spoofed         \\ \midrule
Training                & 8                     & 12                      & 2,580             & 22,800         & 5,400             & 48,600          \\
Development             & 8                     & 12                      & 2,548             & 22,296         & 5,400             & 24,300       \\ \bottomrule  
\end{tabular}%
}
\end{table}

\subsubsection{Data Conditions: Logical Access}
    Logical access includes spoofing attacks generated with text-to-speech (TTS) and voice conversion (VC) attacks. Genuine speech is collected from speakers with no significant channel or background noise effects. Spoofed speech is generated from genuine data using a number of possible spoofing algorithms. The full dataset is split into three subsets, the first one is for training, the second is for development and the third is for evaluation. The number of speakers in the former two subsets is illustrated in Table \ref{tab:dataset}. There is no speaker overlap in these three subsets regarding target speakers used in voice conversion or TTS technique. The voice conversion systems are based on (1) neural-network-based and (2) transfer-function-based methods. The speech synthesis systems were done with (1) waveform concatenation, (2) neural-network-based parametric speech synthesis using source-filter vocoders and (3) neural-network-based parametric speech synthesis using Wavenet.
    
\subsubsection{Data Conditions: Physical Access}
    The spoofing attacks that at the sensor level are considered as physical access scenario. Both bona fide and spoofed signals propagate through a physical space prior to the acquisition. Spoofing attacks in this scenario are therefore referred to as replay attacks, whereby are cording of a bona fide access attempt is first captured, presumably surreptitiously, before being replayed to the ASV microphone.
    
    Table 1 shows the number of speakers and the number of bona fide and replay spoofing access attempts (utterances) in and development set. Both bona fide and replay spoofing access attempts in both training and development data partitions were generated with respect to the same set of randomly selected acoustic and replay configurations.
    
    In a similar fashion to the logical access scenario, the evaluation set is disjoint in terms of speakers. Evaluation data is generated in the same manner as training and development data, albeit with different, randomly acoustic and replay configurations. Specifically, the set of room sizes,  levels of reverberation,  speaker-to-ASV microphone distances,  attacker-to-speaker recording distances, and loudspeaker qualities, while drawn from the same categories, will be different. Full details of the dataset and performance measures of ASVspoof 2019 are presented in evaluation plan \cite{evaluation}.
    %cite evaluation plan

\subsection{Performance measures}
    Improvements in Countermeasure (CM) technology need not necessarily imply an improved complete system that involves co-operation of CM and ASV. Thus, the main focus is on assessment of tandem systems where the CM act as a `mediator' to determine whether a given speech input originates from a bona fide (genuine) user, before passing it as the main biometric verifier (the ASV system). The cost function used in this scenario is a recently-proposed tandem detection cost function (t-DCF) \cite{tdcf} as its primary performance metric. Other than t-DCF, the evaluation of the equal error rate (EER) of each submitted countermeasure is used as a secondary performance metric \cite{evaluation}. %evaluation plan

\subsubsection{Tandem-Detection Cost Function (t-DCF)}
    The t-DCF \cite{tdcf} is mainly consisted of statistical detection theory and involves detailed specification of a specified application. A key feature of t-DCF involves the assessment of a tandem system while on the other hand, keeping the two subsystems (CM and ASV) isolated from each other, i.e. they can be developed independently of each other.
 
    The t-DCF metric has 6 parameters: (i) false alarm and miss costs for both systems, and (ii) prior probabilities of target and spoof trials (with an implied third, non-target prior).
    \begin{equation}
        t-DCF(s) = C_{1}P_{miss}^{cm}(s) + C_{2}P_{fa}^{cm}(s)
    \end{equation}
    where $P_{miss}^{cm}(s)$ and $P_{fa}^{cm}(s)$ are, respectively, the miss rate and the false alarm rate of the CM system at threshold s. The constants C\textsubscript{2} and C\textsubscript{1} are dictated by the t-DCF costs, priors, and the ASV system detection errors.
    
\subsubsection{Equal Error Rate (EER)}
    
    In scoring-based spoofing detection tasks, EER is used as a secondary metric to analyze the performance of different counter measures methods. Let FRR($\theta$) and FAR($\theta$) denote the false rejection rate and false acceptance rate at threshold $\theta$.

    \begin{equation}
        FAR(\theta) = \frac{count(spoof \ trials \ with \ score) > \theta}{total \ spoof \ trials}
    \end{equation}
    \begin{equation}
        FRR(\theta) = \frac{count(human \ trials \ with \ score) < \theta}{total \ human \ trials}
    \end{equation}
    
    $FRR(\theta)$ and $FAR(\theta)$ are monotonically decreasing and increasing functions of $\theta$. The EER corresponds to the threshold $\theta$\textsubscript{EER} at which two detection error rates are equal.

\subsection{Model Setup}
    Our model was trained only on the training set and validated on the development set of the ASVspoof 2019 dataset. The model was trained in Google Colab with the granted GPU (Tesla K80). Audio preprocessing was performed via the Google Cloud Platform. Python's fastai library was used to train CNN which provides support for vision, text, tabular, and collaborative filtering models. The fastai library is a high-level library build on PyTorch which enables easy prototyping and gives you access to a lot of state-of-the-art methods such as marketing application (horizontal application) and health care application (vertical application). Mel-spectrograms were extracted from the audio files using FFT window of size 2048 and the number of samples between successive frames was taken as 512. The Mel-spectrograms were further resized to 224x224 before feeding into the CNN.  ResNet-34 training ran on Google Colab with ADAM optimizer and a learning rate in range of 0.000001, over 8 epochs and a batch size of 64.

\section{Results and Discussion}

\begin{table}[th]
\caption{The experiment result obtained from ASVspoof 2019.}
\label{tab:result}
\centering
\resizebox{\linewidth}{!}{%
\begin{tabular}{@{}ccccc@{}}
\toprule
\multirow{2}{*}{Dataset}                                            & \multicolumn{2}{c}{EER}  & \multicolumn{2}{c}{t-DCF} \\ \cmidrule(l){2-3}\cmidrule(l){4-5} 
                                                                    & Development & Evaluation & Development  & Evaluation \\ \midrule
\begin{tabular}[c]{@{}c@{}}Logical Access\end{tabular}  & 0.9056           & 5.32          & -            & 0.1514          \\
\begin{tabular}[c]{@{}c@{}}Physical Access\end{tabular} & 5.87           & 5.74          & -            & 0.1351       \\ \bottomrule
\end{tabular}%
}
\end{table}

\begin{table}[th]
\caption{The experiment result of baseline in ASVspoof 2019.}
\label{tab:baselineresult}
\centering
\begin{tabular}{@{}ccccc@{}}
\toprule
\multirow{2}{*}{\begin{tabular}[c]{@{}c@{}}Baseline\\ Implementation\end{tabular}} & \multicolumn{2}{c}{Logical Access} & \multicolumn{2}{c}{Physical Access} \\ \cmidrule(l){2-3} \cmidrule(l){4-5} 
                                                                                  & EER             & t-DCF            & EER              & t-DCF            \\ \midrule
CQCC - GMM                                                                        & 9.57            & 0.2366           & 11.04            & 0.2454           \\
LFCC - GMM                                                                        & 8.09            & 0.2116           & 13.54            & 0.3017           \\ \bottomrule
\end{tabular}
\end{table}

    Table \ref{tab:result} shows the result of our model in ASVspoof 2019 challenge based on the two performance matrices (t-DCF and EER) on evaluation and development dataset. Comparing the result of ASVspoof 2019 challenge, our single model outperforms the baseline models provided by ASVspoof 2019 organizers given in Table \ref{tab:baselineresult}. But when compared to other teams, the result that we obtained in the testing phase was low. We had investigated the reason behind this problem and found that in the test dataset, spoofed data are generated according to diverse unseen spoofing algorithms. However, they are variations of the spoofing algorithms used to generate the development dataset.
    
     Among the systems submitted under ASVspoof 2019 challenge, the best-performing system under logical access scenario, had a min-tDCF of 0.2093 and an equal error rate (EER) of 11.40\% (while evaluating a single system). By comparing this with our single system, we had obtained a better result in both the performance metrics. Along with that, while analyzing the result and spoofing attack algorithms provided by ASVspoof 2019 committee, our model performed well in detecting spoofing attacks which uses voice conversion techniques by waveform filtering (A17). In both logical and physical access scenario, we are one among the top three teams while considering both primary and single system as same. We had proposed the same methodology, rather than going for different methodologies to counter both logical and physical access scenarios. Also, t-DCF and EER showed correlation of 0.99686 and 0.96886 in physical and logical access scenarios respectively.
    
    %further research gan & noise
\section{Conclusion}

   In this paper, we explored the applicability of the transfer learning approach for the solution to the problem of spoofing detection. Our first finding is that, instead of feeding raw audio directly to the model, we use an advanced approach to audio classification using Mel-spectrogram. Mel-spectrogram transforms the input raw sequence to a 2-D feature map in which one dimension represents time and the other represents frequency and the values represent amplitude. A second finding is that residual neural networks can help us build deeper models that outperform conventional deep neural networks. 
   
   Finally, we demonstrated that even with a single model, better system performance can be obtained. By using different features and models system performance can be significantly improved. In future, we would like to explore the effectiveness of using more advanced generative adversarial networks (GAN) for the spoofing detection task. Denoising methods like Adaptive filtering and Autoencoders could also be implemented.  

\section{Acknowledgements}
    The authors would like to thank ASVspoof 2019 organizers for providing the dataset and detailed analysis of our system. We also acknowledge the technical support from Sachin Dev S, working at Tricodia Inc.

\bibliographystyle{IEEEtran}

%\bibliography{mybib}

\end{document}